# The "0.4 eV" Shape Resonance of Electron Scattering from Mercury in a Franck-Hertz Tube

**Peter Nicoletopoulos**[1]

*Faculté des Sciences, Université Libre de Bruxelles, Brussels, Belgium*

E-mail: pnicolet@skynet.be and pnicolet@ulb.ac.be

**Abstract**

The alternative version of the Franck-Hertz experiment with mercury, in which a two-grid tube is used as a combination of electron gun, equipotential collision space, and detection cell, was analyzed recently in considerable detail. In particular, it was inferred that, at optimal pressure, the formation of peaks in the anode current at inelastic thresholds is mediated inside the detection cell by the large variation, a maximum at 0.4 eV, in the cross section for *elastic* scattering. This variation is due to a shape resonance in the electron-mercury system and is observable persuasively at the onset of anode current as a sharp peak followed by a clear minimum. In the present paper, the passage of electrons through the second grid to anode region is analyzed in terms of kinetic theory. The discussion is based on a simplified expression for the electron current derivable from an approximate form of the Boltzmann transport equation that maintains the spatial density gradient but omits elastic energy losses. The estimated range of pressure underlying this kind of idealization is in good agreement with experiment. An explicit solution is obtained by constructing an analytic expression for the momentum transfer cross section of mercury using a recent theory of generalized Fano profiles for overlapping resonances. This solution is used in order to model successfully the formation of peaks at the threshold of anode current and at excitation potentials, and to explain the dependence of the observed profiles on the pressure and on the sign and magnitude of the potential across the detection cell.

---

[1] Address for correspondence: Rue Joseph Cuylits 16, 1180 Brussels, Belgium.

## 1. Introduction

The standard Franck-Hertz experiment and the alternative arrangement designed to show more than one excited state, are extraordinarily elegant demonstrations of quantum mechanics that are deceptive in their simplicity.

In the latter version, electrons are accelerated in less than one excitation mean-free-path to a final potential and allowed to travel across a field-free region between two grids. A recent analysis of extensive data obtained in this experiment with mercury [1] revealed that in order to unravel the results, (a) several complexities due to space charge must be understood and (b) the effect of *elastic* collisions on electrons traversing the detection cell (the region between $g_2$ and A, in Figure 1), must be taken into account.

In the first place, the potential inside the drift cell between the two grids ($g_1$ and $g_2$) is modified dynamically by the interaction of negative and positive space charges. This complicates the interpretation of the observed excitation spectra because some peaks can be due to the same feature excited in different regions of the tube.

A determination of the space potential can be obtained from a careful examination of the change in the spectra induced under controlled conditions. It was inferred that if a *decelerating* voltage ($\Delta V$) is applied across the main collision cell, this region is separated into two field-free, weakly ionized plasma regions at different potentials, separated by a narrow space-charge double layer (see Fig. 1). Varying the intergrid potential can change the position of the double layer. One can thus manage to create a field-free region of substantial width at the potential of the exit grid. This space constitutes the useful scattering chamber. A good quality spectrum is shown in Fig. 2.

The second major point is that, in the optimal range of pressure, the motion of electrons at low electric field is governed primarily by diffusion. As a result, the selection of electrons that have undergone inelastic collisions is mediated inside the detection cell by a large variation (a maximum at about 0.4 eV) in the *elastic* cross section of electron-mercury scattering.

This aspect of the experiment is the main theme of the present paper.

Apart from its implicit appearance in the profiles of excitation peaks, the variation in the elastic cross section is also observable directly: a peak is seen close to the onset of anode current followed by a minimum (see feature 1 in Fig. 2). The demonstration of this feature in its own right can be planned as a separate experiment suitable for a university laboratory. It should be feasible with any Franck-Hertz tube and would serve to show that elastic scattering too can be strongly energy-dependent as originally pointed out by R. Minkowski [2], who originally discovered this phenomenon in mercury and cadmium with a two-grid Franck-Hertz tube 80 years ago.

Better known variations at low energy are due to a different quantum mechanical mechanism–the Ramsauer-Townsend effect–which is observed in the elastic cross-sections of the heavier noble gases (Ar, Kr, Xe, Rn). This phenomenon should be particularly easy to demonstrate with the present method, as outlined in Sec. 5.

## 2. Drift and diffusion without energy loss.

A theory of electron transport suitable to the present context is based on the following continuity equation for the flux density $\Gamma$ under a constant electric field

$$\Gamma = \frac{1}{3}\frac{eE\lambda_M(x)}{mv(x)}n(x) - \frac{1}{3}v(x)\lambda_M(x)\frac{dn(x)}{dx} \qquad (1)$$

where x is the space coordinate along the electric field, n(x) is the electron number density, m, e, the electron mass and charge, E the electric field strength, v(x) the electron thermal velocity and $\lambda_M(x)$ the mean free path for momentum transfer. The first term in (1) represents the current carried by drift and the second the current carried by diffusion.

This equation was given long ago by Hertz [3]. The main assumption is that the mass M of neutral atoms is infinite and hence the electron energy $\epsilon$ is conserved: $\epsilon(x)=(1/2)mv^2(x)=\epsilon(x_0)+eE\times(x-x_0)$. Thus only the velocity vector **v** is affected by collisions.

A modern study by McMahon [4] shows that, in infinite parallel-plane geometry, equation (1) follows easily from the Boltzmann kinetic equation for the electron distribution function f(x,v), using the common two-term expansion

f(x,v)=f₀(x,v)+f₁(x,v)cosθ, with **v.E**=-vEcosθ, and letting M→ ∞ before proceeding to the solution.

This simplified theory should be applicable between two electrodes if their spacing d is less than the mean free path for energy transfer $\lambda_\varepsilon$, where $\lambda_\varepsilon = \lambda_M (M/2m)^{1/2}$. Since d must also be much larger than $\lambda_M$ for the Boltzmann equation to hold, the requirements for validity can be stated as

$$\lambda_M \ll d < \lambda_\varepsilon, \text{ namely, for mercury: } \lambda_M \ll d < 429 \lambda_M \qquad (2)$$

The consistency of the bounds (2) with the experimental conditions delineated in [1] will be verified later on. Considerations relating to electrode geometry are not significant [4-5], and can be disregarded in a semi-quantitative picture.

Integration of Eq. (1) with appropriate boundary conditions leads to an analytic expression for the ratio $I_A/I_S$ of injected versus incident current between two electrodes. The derivation is given in [4] and can also be found in [5]. For monoenergetic electrons starting isotropically with energy $\varepsilon$ from a plane situated at $g_2$ and reaching the anode with energy $\varepsilon_A$ after being accelerated by a constant electric field E, [eE=($\varepsilon_A$-$\varepsilon$)/d] the result is:

$$I_A = I_S(\varepsilon)\left[1 + \frac{\gamma\varepsilon}{\alpha\varepsilon_A} + \frac{3\gamma d}{4}J(\varepsilon)\right]^{-1} \qquad (3)$$

where γ and α are the respective absorption coefficients of the grid and anode; these are assumed to be energy independent and their value can be as large as two; d is the $g_2$-anode distance and J is given by:

$$J = \frac{\varepsilon}{\varepsilon_A - \varepsilon} \int_\varepsilon^{\varepsilon_A} \frac{dy}{y\lambda_M(y)} \qquad (4)$$

The portion of $I_S$ absorbed by the grid is given by

$$I_G = I_A(\varepsilon)\left[\frac{\gamma\varepsilon}{\alpha\varepsilon_A} + \frac{3\gamma d}{4}J\right] \qquad (5)$$

so that $I_S=I_A+I_G$. This treatment improves on a previous solution by Langmuir [6] since it includes the full dependence on a velocity-dependent free path. For pure diffusion the function J tends to $\lambda_M^{-1}(\varepsilon)$ (since $\varepsilon_A \to \varepsilon$ as $E \to 0$). In such conditions, and at the larger pressures, equations (3) and (5) reduce to

$$I_A = I_S(\varepsilon)\frac{4}{3\gamma d}\lambda_M(\varepsilon) \tag{6}$$

$$I_G = I_S(\varepsilon)\left(1 - \frac{4}{3\gamma d}\lambda_M(\varepsilon)\right) \tag{7}$$

Clearly, the grid picks up most of the incident current.

In order to apply the theory to the experiment described in [1] we need an expression for the momentum transfer cross section of mercury.

## 3. An analytic expression of $\sigma_M(\varepsilon)$ for mercury.

The total cross section, $\sigma_T(\varepsilon)$, and the momentum transfer cross section, $\sigma_M(\varepsilon)$, for elastic electron scattering from mercury (and several other elements) exhibit a large peak at low energy that is attributed to an unstable state (a negative ion). A recent review of this subject is given in [7].

The asymmetric nature of these peaks suggests that these cross sections should be expressible analytically in terms of the well-known Fano lineshape [8] for an isolated resonance (the lowest unstable state of the negative ion) in a "flat" continuum (decay channel). But a careful comparison with experimental profiles shows that this is impossible. Interference effects caused by broad higher-energy negative ion resonances imbedded in the same continuum significantly modify the single-resonance Fano lineshape. An extensive analysis is given elsewhere [9].

Recently, a generalized expression of cross section profiles involving interfering resonances was developed in [10-11]. This lineshape is simply a product of Fano factors and of a Breit-Wigner term representing the portion of the continuum (lumped essentially into a broad "quasi-bound state", a wave packet) that is strongly coupled to the resonances. For two resonances, the

generalized lineshape has the form

$$F(\varepsilon) \approx \frac{(q_1+e_1)^2}{1+e_1^2} \frac{(q_2+e_2)^2}{1+e_2^2} \frac{1}{1+e_3^2} \tag{8}$$

where $e_i=2(\varepsilon-\varepsilon_i)/\Gamma_i$, with $\varepsilon_i$ the resonance energies and $\Gamma_i$ their width, and the $q_i$ are constants. It was found [9], that expressions of the form $AF(\varepsilon)+B$, with suitably adjusted constants A, B, provide analytic expressions for resonance-dominated elastic cross sections of several elements, that are quite accurate in a wide energy range. The constant B is viewed as a phenomenological quantity representing dynamics involving states that are not included in the model subspace spanned by the resonances, on which $F(\varepsilon)$ is based.

Figure 3 shows the momentum transfer cross section of mercury derived in [12] from inversion of drift velocity data in a swarm experiment, in comparison to an analytic $\sigma_M(\varepsilon)$ of the form $AF(\varepsilon)+B$. The latter was constructed in [9] in terms of Eq. (8) with only one resonance factor, using the parameters

$$q_1=0.61,\ \varepsilon_1=0.145,\ \Gamma_1=0.39,\ \varepsilon_2=0.34,\ \Gamma_2=1.65,\ A=442.5,\ B=91 \tag{9}$$

where energies are given in eV and A, B in squared Bohr radii ($a_0^2$).

The maximum of our analytic cross section is at $\varepsilon=0.42$ eV in accordance with the value determined in an accurate beam experiment [13]. The idealized energy of the resonance, however, is much lower, as indicated by some theoretical calculations [7]. This property is characteristic of single-resonance profiles with a shape parameter $q<1$ and is the reason for the quotation marks in the title of this paper. Only for $q>>1$ does the resonance energy coincide with the maximum in the cross section.

## 4. Peak formation by the resonance

For the mean free path $\lambda_M(\varepsilon)$ (cm) in terms of $\sigma_M(\varepsilon)$ ($a_0^2$) and of the vapor pressure p (Torr) at temperature T (°C) we use the convenient expression (accurate within 1%):

$$\lambda_M(\varepsilon) = [N\sigma_M(\varepsilon)]^{-1},\ \text{with}\ \ N = p\frac{273}{273+T}$$

In an energy range encompassing the peak in $\sigma_M(\varepsilon)$, the interval of vapor pressure p imposed by condition (2) using d=0.2 cm is between about 0.1 and 6 Torr (80°<T<170°). This range is indeed between the minimum required for observing feature 1 and the maximum allowed for satisfactory display of excitation peaks (see Sec. 8.1 in [1]). Below this interval there is no diffusion; above this range the experimental results are degraded, presumably because energy losses due to elastic collisions become significant with resultant breakdown of the simple Hertz-McMahon picture.

In order to model actual observations, we must include the effect of the electron energy distribution. In the usual manner [14], we let $I_S(\varepsilon_m)f(\varepsilon-\varepsilon_m)d\varepsilon$ represent the energy distribution of the current at detection-cell pressure zero where

$$\int f(\varepsilon - \varepsilon_m)d\varepsilon = 1$$

and $I_S(\varepsilon_m)$ is the total current to the anode at mean energy $\varepsilon_m$, (namely at applied potential $V=\varepsilon_m/e$ ). When the energy distribution is folded into Eq. (3), we obtain

$$I_A(\varepsilon_m) = I_S(\varepsilon_m)\int_0^\infty f(\varepsilon-\varepsilon_m)\times\left[1+\frac{\gamma\varepsilon}{\alpha(\varepsilon+F)}+\frac{d\gamma\varepsilon}{F}\int_\varepsilon^{\varepsilon+F}\frac{N\sigma_M(y)}{y}dy\right]^{-1}d\varepsilon \quad (10)$$

where F/e (namely $|\Delta a|$ in Fig. 1) is the accelerating potential between the second grid and the anode. If a decelerating potential is applied ($\Delta a<0$), the expression inside the bracket in (8) must be replaced by

$$1+\frac{\gamma(\varepsilon+F)}{\alpha\varepsilon}-\frac{d\gamma(\varepsilon+F)}{F}\int_{\varepsilon+F}^\varepsilon\frac{N\sigma_M(y)}{y}dy \quad (11)$$

and it should be understood that in this case the threshold ($\varepsilon_m=0$) of anode current is displaced by F.

We first consider the mechanism of formation of feature 1 in Fig. 2.

The energy distribution of cathode electrons arriving at the second grid at various pressures is difficult to evaluate. The abrupt onset of the anode current suggests a short-tailed form. The results shown below were calculated with a Gaussian distribution $f(\varepsilon)=(\pi\Omega^2)^{-1/2}\exp[-\varepsilon^2/\Omega^2]$ with $\Omega=0.2$ eV. Similar profiles are obtained using a Maxwellian whose width is close to the estimated temperature ($T_c$) of the cathode namely $f(\varepsilon)=(2\Omega)^{-1}\exp[-(\text{sgn}\varepsilon)\times\varepsilon/\Omega]$, with $\Omega=0.1$ eV, ($T_c\sim 1000°$ K).

Calculating $I_A(\varepsilon_m)$ involves a simple sequence of numerical integrals, because with the analytic cross section of Sec. 3 the integral inside the bracket can be evaluated in closed form.

A first approximation to the current profile near threshold can be obtained by using Eq. (10) with $I_S(\varepsilon_m)=1$. We take p=4.2 Torr (T=160°) d=0.2, $\alpha=1$, $\gamma=0.1$; this value of $\gamma$ for the type of tube employed in [1] was estimated by McMahon in [5].

Figure 4 shows the results for a positive $\Delta a$ (F= 0.3 V), for the field-free case F=0, and for retarding potentials of 0.2, 0.5 and 4 volts. It is seen explicitly that for negative values of $\Delta a$ the variation in the current is smoothed out: *Feature 1 is not observable under retarding potential conditions.* This is why the peak near the onset of anode current does not appear in the current-voltage curves of the earlier authors referenced in [1]. Minkowki, however, did use a positive $\Delta a$ in his original demonstration of this structure [2].

The next step is to model the current at different pressures (keeping $I_S=1$). The resulting profiles are shown in the dotted curves of Fig. 5. More realistic curves are obtainable by including the influence of a rising injected current. The current $I_S$ can be written as $I_S=I_0(\varepsilon_m)\times I_T(\varepsilon_m)$; $I_0$ is the fraction of cathode current injected at the first grid with energy $e|\Delta V|+\varepsilon_m$, and $I_T$ is the fraction of $I_0$ that reaches the vicinity of the second grid with energy $\varepsilon_m$, after traversing the large decelerating region between the grids. The factor $I_T(\varepsilon_m)$ varies slowly with $\varepsilon_m$, but $I_0$ is inevitably space charge limited and is thus a rapidly rising function which can considerably distort the observed profile. The effect is illustrated in the full curves of Fig. 5 computed with a linearly rising source current $I_S(\varepsilon_m)=0.4+\varepsilon_m$. These profiles are quite similar to those commonly observed without particular precautions in any tube.

In an interval of large accelerating potential ($V_{g_1}$ in Fig. 1), the rate of rise of $I_0(\varepsilon_m)$ is smaller and hence the distortion of the anode current in the vicinity of the resonance can be reduced by increasing $|\Delta V|$. This effect is evident in the series of profiles of feature 1 displayed in [1].

Basically similar considerations apply to the formation of peaks due to excitations incurred in the *lower-potential* drift space of Fig. 1 (as shown in [1], those are the peaks beyond feature 1 labeled by numbers in Fig. 2). Since that collision-cell is at $g_2$ potential, a source of slow electrons will be formed at the entrance of the $g_2$-anode region at every inelastic threshold $\varepsilon_m=\varepsilon^*$. It is presumed that only an insignificant number of inelastic events take place inside the detection-cell, an assumption that is supported by visual observations, as described elsewhere [15].

It follows that the previous equations are applicable for modelling qualitatively the formation of excitation peaks with the understanding that energies are measured with respect to $\varepsilon^*$.

Note that beyond the first inelastic threshold the cathode current has reached saturation, and so the distortion due to a rising source current does not take place. On the other hand, $I_S(\varepsilon_m)f(\varepsilon-\varepsilon_m)d\varepsilon$ will now depend on the particular form of the inelastic cross section, on the pressure, and on the length of the effective collision chamber. A detailed analysis of the shape and strength of individual excitation peaks is beyond the scope of the present paper.

The influence on excitation peaks of the sign of $\Delta a$ is not easy to verify experimentally. Small changes of this potential in either direction affect primarily the background current (a smoothly varying combination of fast electrons and positive ions) so that the whole excitation curve is shifted inconveniently upwards or downwards. All one can say is that $\Delta a=0$ was found to be the best setting, except for some cases where a small positive value of 0.1-0.3 V was necessary in order to obtain a well-balanced curve throughout the energy range.

## 5. Employing the method for showing the Ramsauer effect

The clarification of Minkowski's experiment obtained here in terms of kinetic theory suggests that this type of apparatus would be particularly effective for demonstrating the Ramsauer-Townsend effect in the noble gases. A large peak in the anode current should be observed at low energy that coincides with the deep minimum in the elastic cross section.

The anticipated result can be modelled with the tools presented here. An analytic momentum transfer cross section typical of the heavier noble gases, with a minimum at 1 eV and a maximum close to 10 eV, can be constructed using the generalized Fano formula (8) of Sec. 3 with a fictitious resonance of negative shape index q.

The values of the parameters are

$$q_1=-0.9, \varepsilon_1=0.1, \Gamma_1=2, \varepsilon_2=8, \Gamma_2=15, A=200, B=1 \qquad (12)$$

and the cross section is shown in Fig. 6.

Assuming that the gas is xenon [$(M/2m)^{1/2}=358$], the large-pressure limit implied by condition (2) of Sec. 2 is now about p=11 Torr at T=25° (N=10). Figure 7 shows the anode current obtained at this pressure from Eq. (10), for F=0 and with the previous tube parameters. The dashed curve is for $I_S(\varepsilon_m)=1$ and the solid curve shows a more realistic form, resulting from a linearly rising source current $I_0(\varepsilon_m)=0.4+\varepsilon_m$.

A demonstration of the Ramsauer effect in xenon has been in the literature for some time [16]. But that experiment, carried out in a commercial 2D21 xenon thyratron at about 0.05 Torr, relies on a beam type technique (no diffusion): the anode current is simply a measure of the unscattered portion of the beam. Significant discrimination between the scattered and unscattered components occurs because in that tube the size of the exit aperture towards the collector is limited. Hence $I_A(\varepsilon)/I_S(\varepsilon) \sim \exp[-A\sigma_T(\varepsilon)]$ where $\sigma_T$ is the total cross section and A=Nx, x being the length of the collision cell. For a sufficiently large value of A, this form of current too exhibits a large maximum at the minimum of $\sigma_T(\varepsilon)$. So to some extent the profile of $I_A(\varepsilon)$ obtained in [16], is similar to that of Fig. 7 although the latter results from $I_A(\varepsilon)/I_S(\varepsilon) \sim 1/\sigma_M(\varepsilon)$.

Contrasting the beam-type demonstration with the observation of the same phenomenon in the *diffusion* current in a Franck-Hertz tube (namely in *open* collector geometry) would be an instructive introduction to the relationship of fundamental atomic processes to the macroscopically observable transport properties of a gas. The design and construction of a suitable tube should be straightforward.

## 6. Conclusion

The main lesson conveyed in this paper is that the remarkable results of the extended Franck-Hertz experiment can only be appreciated in terms of kinetic theory. The description in terms of simple beam attenuation commonly found in textbooks is not pertinent because it is inconsistent with the range of pressure required for producing acceptable results. For a beam picture to be operative the gas temperature must be below 80°. Structure is in fact descernible below this temperature (as mentioned in Sec. 8.1 of [1]) but its definition is far from satisfactory. An amended arrangement incorporating a sophisticated detection mechanism must be used in the low-pressure regime, as described in the work of Martin and Quinn [17]. And at any rate, in the lack of significant diffusion the observation of the 0.4 eV resonance is impossible because no discrimination between the scattered and unscattered portions of an elastically colliding beam is provided for.

Similar remarks apply to the textbook explanation of the *standard* Franck-Hertz experiment. That a truer picture highlighting electron transport effects is long overdue, is easily seen simply by verifying that the retarding potential between $g_2$ and anode–a mandatory accessory in the beam interpretation– is beside the point, as pointed out by McMahon [5].

Actually the standard version is even more difficult to elucidate. The pressure recommended in this case with the tube of [1] (T=200°-220°, namely p=17-32 Torr) exceeds the range of the Hertz-McMahon regime for mercury so that the full weight of modern kinetic theory must likely be brought to bear. Work in this direction was reported recently by Robson *et al* [18-19], and Sigeneger *et al* [20]. Still, their approach is only a preliminary step because (a) the role played by the detection cell between grid and anode is not explained and (b) the recurrent variations of potential observed in striated discharges [21] (of the type underlying the standard Franck-Hertz experiment [22]) are disregarded.

The mechanisms leading to field-variations in striated discharges are currently being investigated by Goluboskii *et al*. A glance at [23] is helpful for appreciating the complexity of the problem.

**Figure captions**

Fig. 1.  Interelectrode space-potential between cathode (K) and anode (A) showing two field-free plasma regions joined by a space-charge sheath. The high-potential plasma is initially (solid curve) at applied value of grid 1 voltage ($V_{g_1}$), but is raised dynamically (dashed curve) by $\delta$ volts in the course of the experiment (see [1] for details). For sufficiently large intergrid potential $\Delta V+\delta$, the high potential region shrinks so that only the spectrum excited at $g_2$ potential is observed.

Fig. 2.  Anode current plot with $\Delta V=-4.1$ V showing a pure low-potential excitation spectrum plus feature 1 (structures labelled by numbers). The early (lettered) peaks are excited at $V_{g_1}$ potential (the low-potential plasma has not been created as yet).

Fig. 3.  The analytic expression of $\sigma_M(\varepsilon)$ for mercury obtained from Eqs (8) and (9) (solid curve) compared to the momentum transfer cross section of England and Elford [12] (dashed curve).

Fig. 4.  Anode current profiles calculated with $I_S=1$, at p=4.2 Torr (T=160°), for grid to anode potentials $\Delta a=+0.3$ V (upper curve) followed in descending order by $\Delta a=0$, $-0.2$, $-0.5$, $-4$. For large negative $\Delta a$ (dashed curve) the rise is almost linear. For easier comparison, the thresholds of the $\Delta a<0$ curves have been shifted leftward to zero.

Fig. 5.  Profiles (with $\Delta a=0$) of the anode current at T=80°, 120°, 160°, corresponding to vapor pressures p of 0.09, 0.75, 4.2 Torr, in descending order. Dashed curves are for $I_S=1$, solid curves for $I_S=0.4+\varepsilon_m$.

Fig. 6.  Approximate analytic momentum transfer cross section for a heavy noble gas, constructed from Eqs. (8) and (12).

Fig. 7.  Predicted Franck-Hertz current profiles demonstrating the Ramsauer-Townsend effect, calculated using the cross section of Fig. 6, at p=11 Torr, T=25°, $\Delta a=0$, with the Gaussian energy distribution and the tube parameters of Sec. 4. Dashed curve is for $I_S=1$, solid curve for $I_S=0.4+\varepsilon_m$.

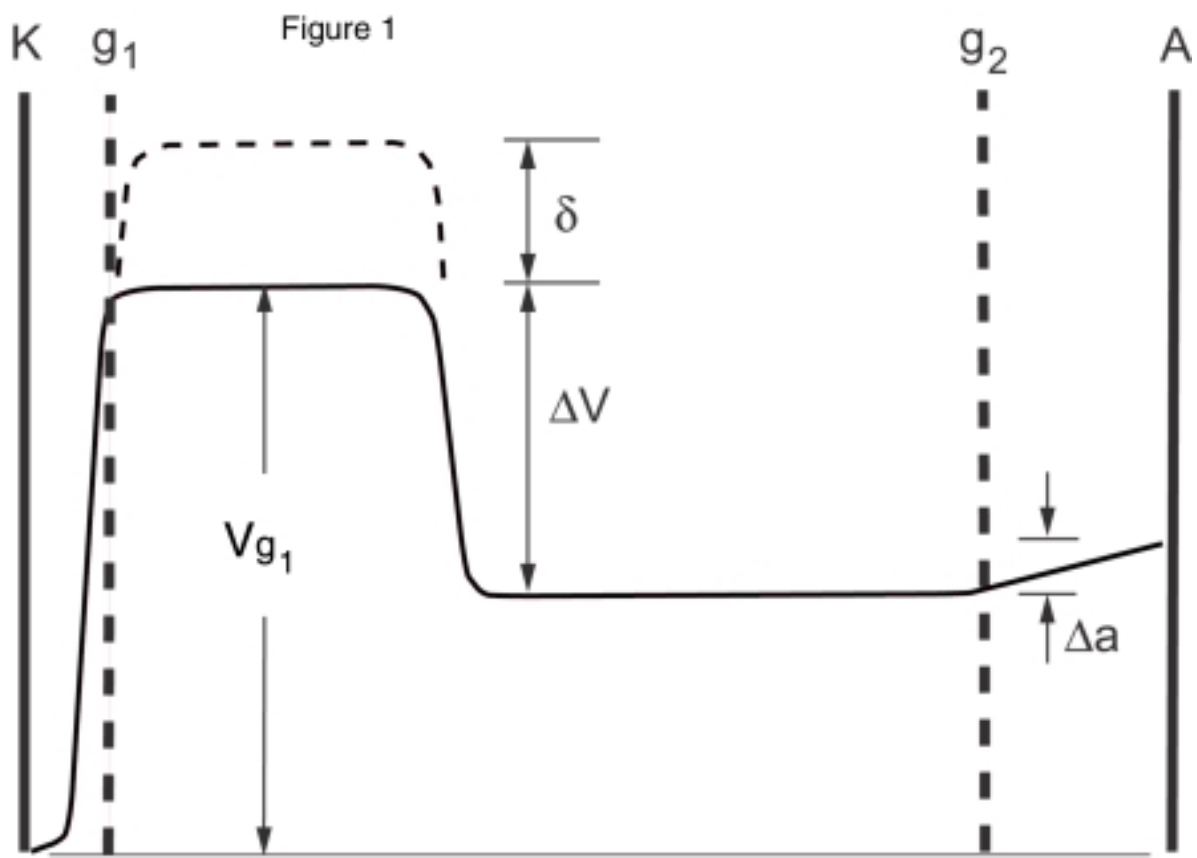

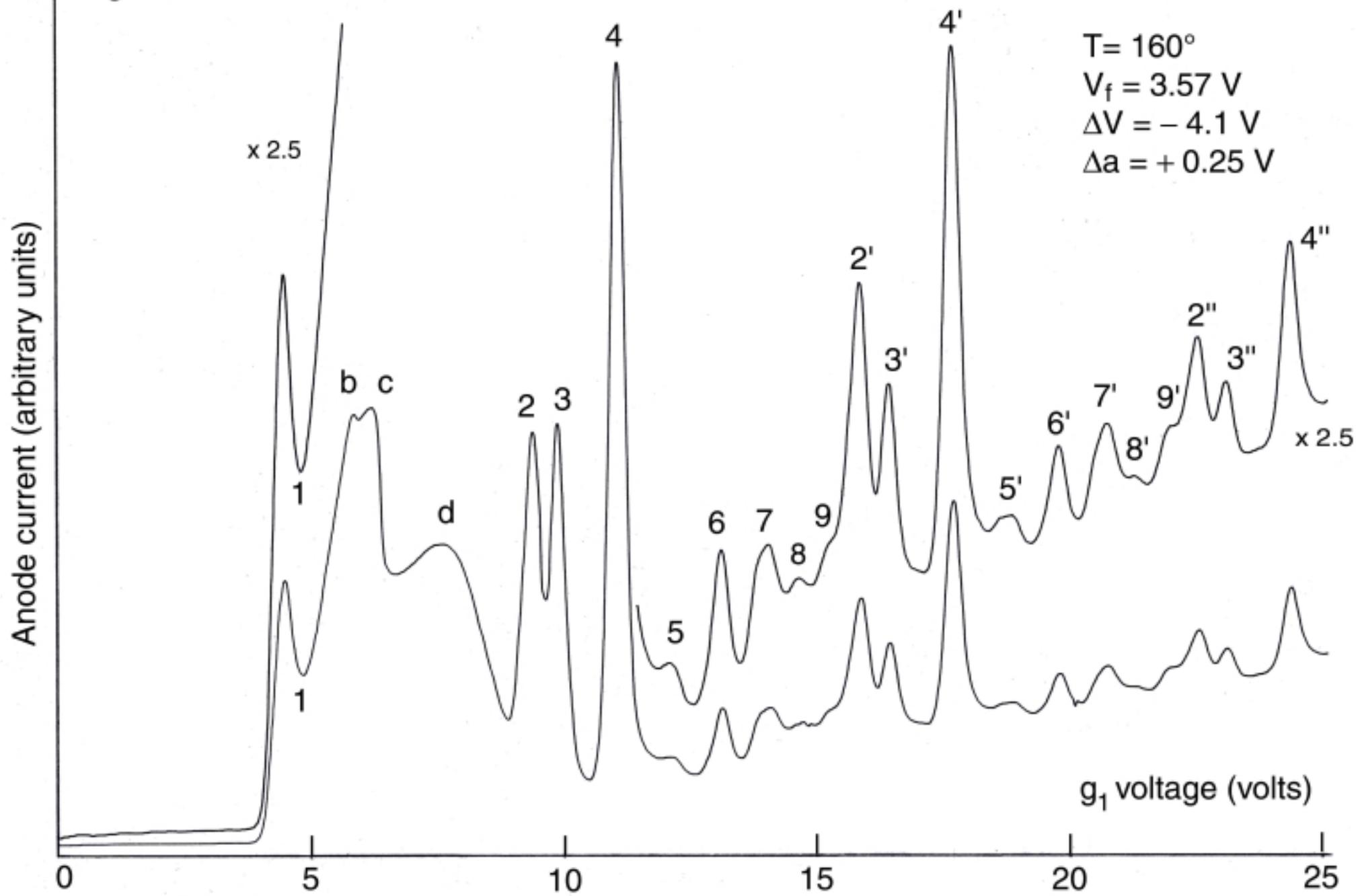

Figure 2 — Sept. 24, 1986

T = 160°  
$V_f = 3.57$ V  
$\Delta V = -4.1$ V  
$\Delta a = +0.25$ V

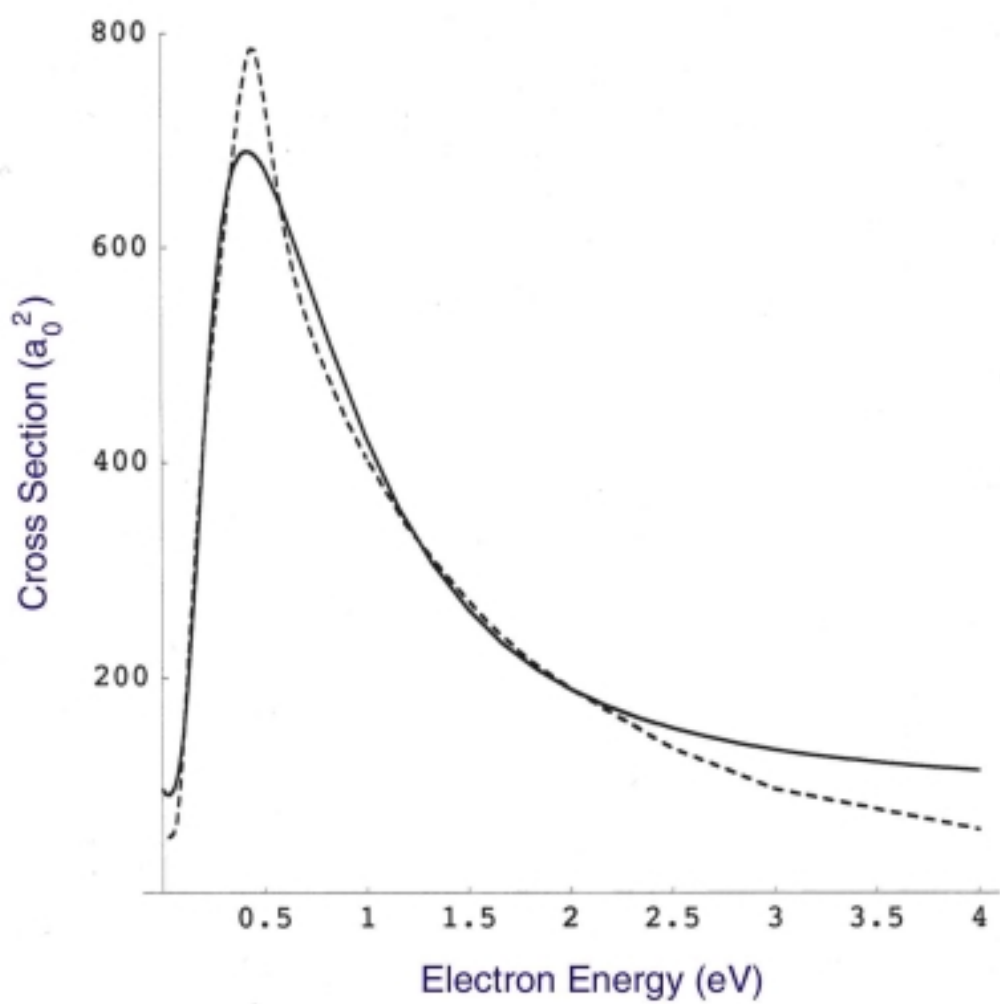

Figure 3

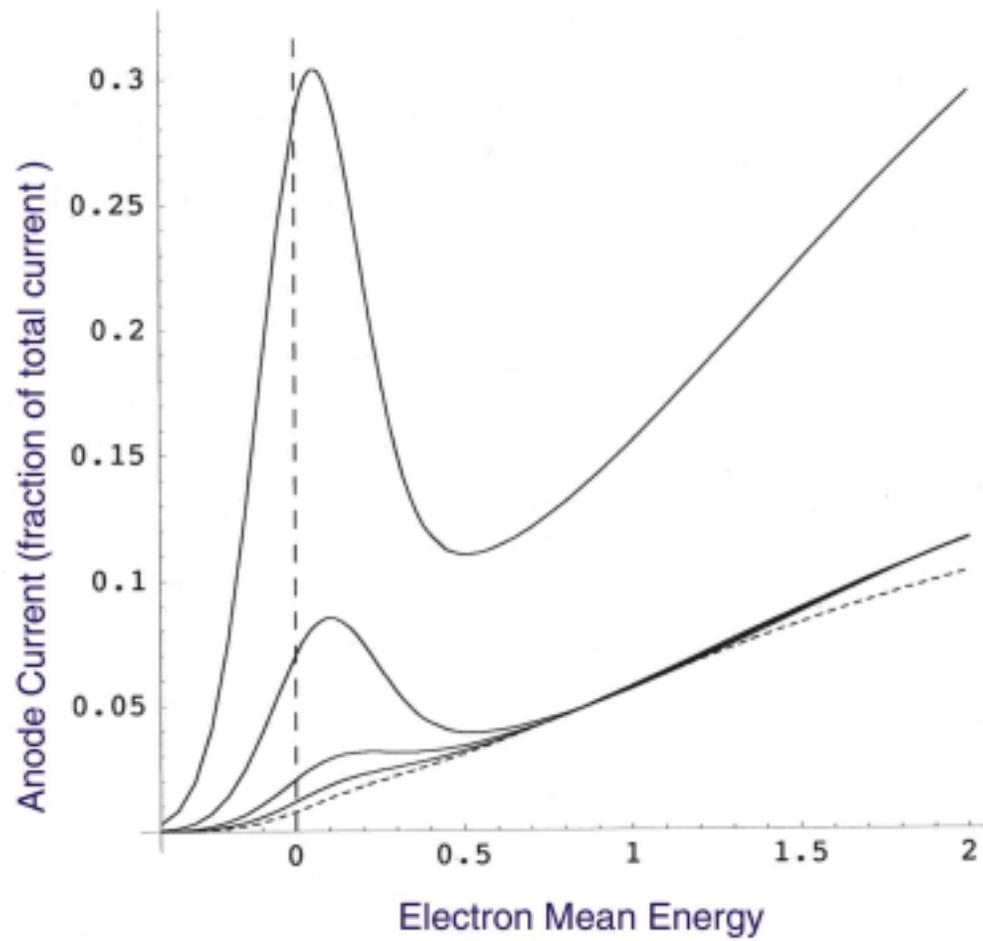

Figure 4

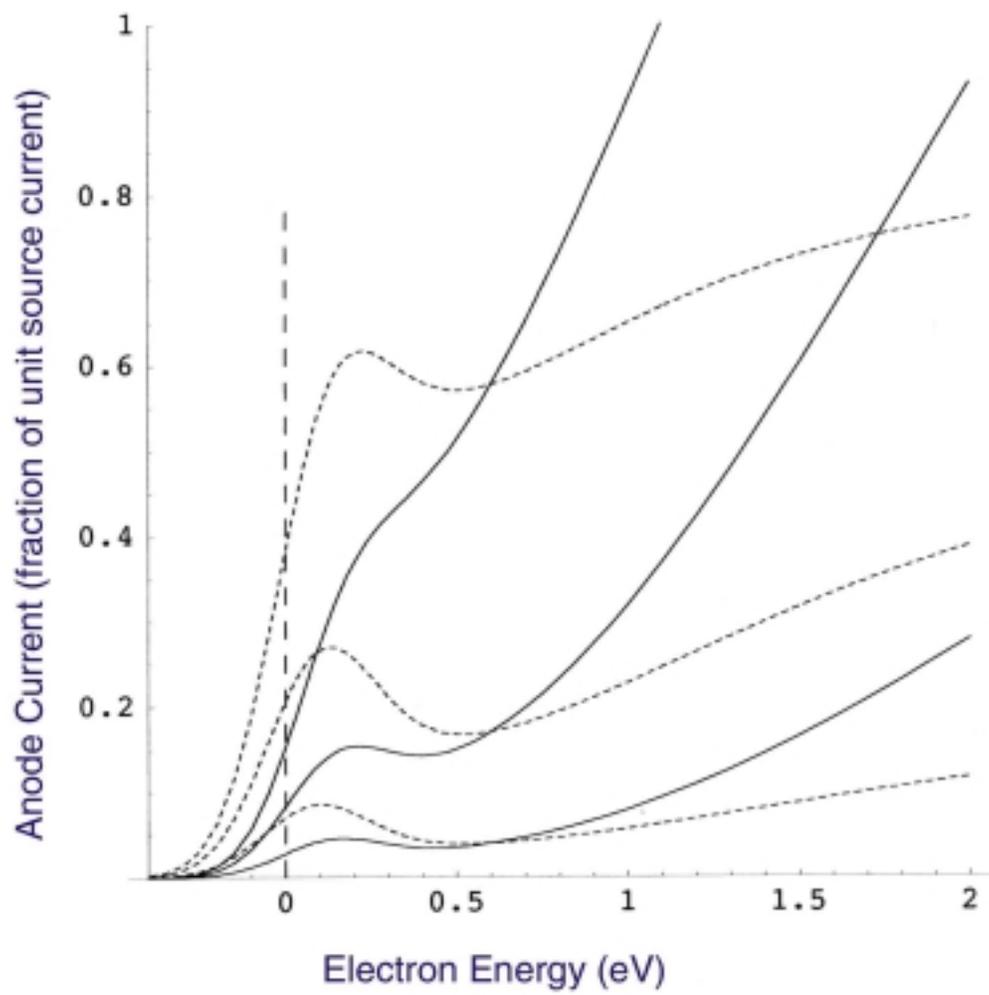

Figure 5

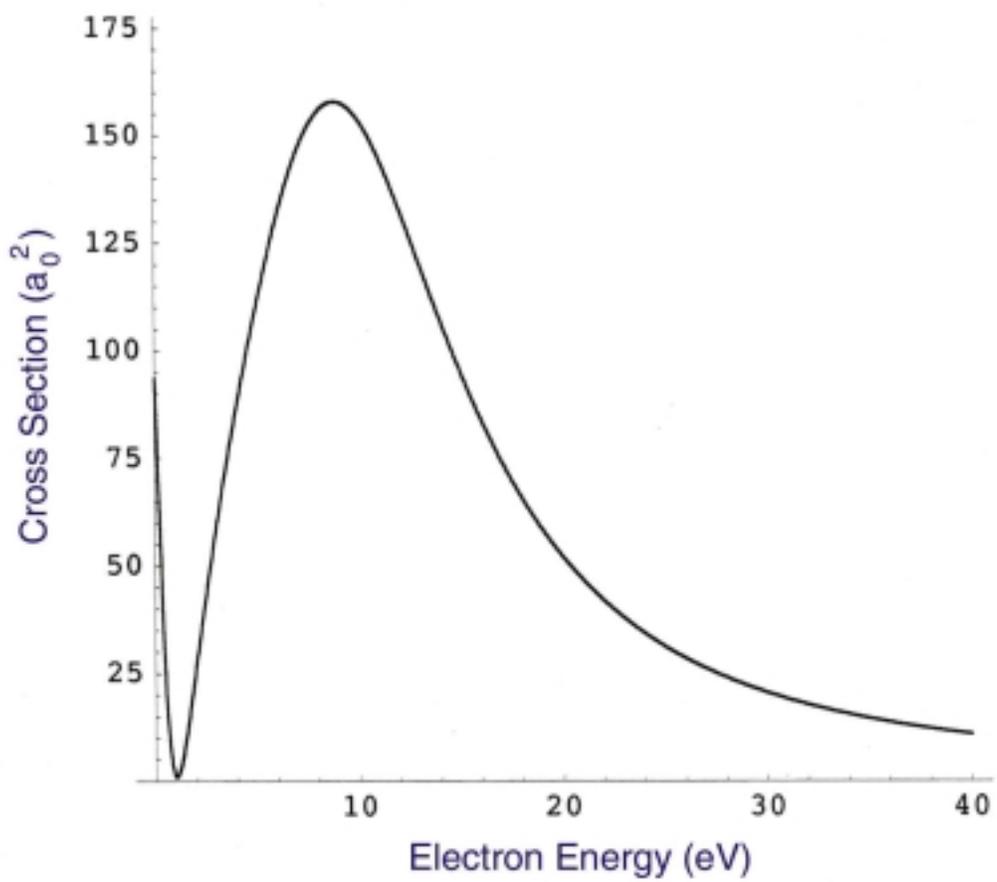

Figure 6

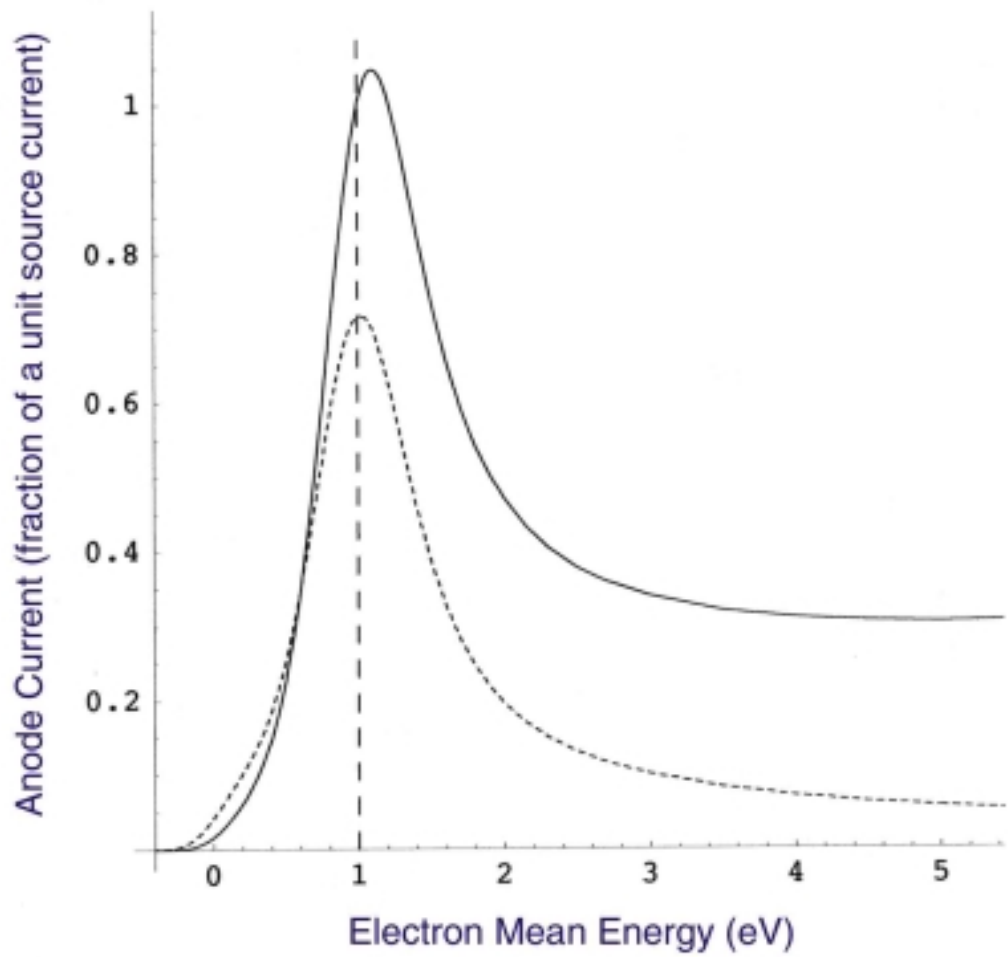

Figure 7